\documentclass[11pt,a4paper]{article}
\usepackage[utf8]{inputenc}
\usepackage{amsmath}
\usepackage{amsfonts}
\usepackage{amssymb}
\usepackage{graphicx}
\usepackage{bm}
\usepackage{booktabs}       
\usepackage{multirow}
\usepackage{epstopdf}
\usepackage{hyperref}       
\usepackage{url}            
\usepackage{bm}             
\usepackage[top=2.52cm, bottom=2.52cm, left=2.52cm, right=2.52cm]{geometry}
\usepackage{xcolor} 
\usepackage{setspace}
\begin{document}
\title{\textbf{The Nonreciprocal Mie-surfaces}}



\author{Dheeraj Pratap\footnote{Email: dpratap@iitd.ac.in}\\ \\
	Optics and Photonic Centre, Indian Institute of Technology Delhi, Delhi 110016, India\\
	}

\date{}

\maketitle

\begin{abstract}
Hemispherical amorphous silicon nanoparticles exhibit asymmetric optical scattering for forward illumination (base-to-apex) and backward illumination (apex-to-base). There exists an anapole mode only for backward propagation, not for forward. Due to the anapole, light is allowed to scatter maximally along the forward direction, and not in the backward direction. A structured surface obtained by repeating hemispheres in a square grid in air exhibits nonreciprocal reflection and transmission for light propagating through it. This nonreciprocity only depends on the diameter of the hemisphere, not on the periodicity. The same surface on a glass substrate causes a minor spectral redshift in the nonreciprocity. Here, the individual materials are Lorentz reciprocal, but the current nonreciprocity is due to interference. The current nonreciprocity is purely based on anapole of Mie scattering; therefore, the surface is termed as ``Nonreciprocal Mie-surface''. Such surfaces could be used for the applications of passive linear nonreciprocal photonic devices.

\vspace{3mm}
\end{abstract}


\section{Main}
The ideas of optical reciprocity and nonreciprocity are fundamental to the study of light behaviour, particularly in the design of optical materials, devices, and systems. The behaviour of light in one direction compared to the opposing direction is described by these two phenomena. The principle that light in an optical system behaves symmetrically when its direction is reversed is known as optical reciprocity~\cite{potton2004reciprocity}. This means that if light can traverse from one point to another point, it will travel from the latter point to the former point in the same manner, provided that the medium and surroundings remain unchanged. Refraction, diffraction, mode conversion, and polarization conversion are among the optical processes that most closely adhere to reciprocity. Unlike reciprocity, optical nonreciprocity describes a condition in which the direction of light transmission affects its behaviour~\cite{manipatruni2009optical}. This implies that the way light moves from the first point to the second point can differ from that of light moving from the second point to the first point. In order to travel or focus light in particular directions and avoid feedback or interference, nonreciprocal optical phenomena are essential. Certain materials or devices that disrupt the symmetry of light transmission exhibit nonreciprocal behaviour and have asymmetric electric or magnetic properties~\cite{poddar2022automatic}. Nonreciprocity can be used to realize one-way light propagation, as in optical circulators and isolators~\cite{Nagulu2020nonreciprocal}.  Breaking time-reversal symmetry is the key to nonreciprocity and can be caused by external sources such as magnetic fields or manufactured materials that react asymmetrically to light. Materials with asymmetric permittivity or permeability tensors\cite{wang2009observation}, time-varying systems~\cite{estep2014magnetic}, and nonlinear light–matter interactions~\cite{shi2015limitations} are the three possible routes for breaching optical reciprocity. In all these approaches, the methods are either externally biased. The innovative properties of metasurfaces, the engineered surfaces with subwavelength features that manipulate light in novel ways, combine with the principles of optical nonreciprocity to create an optical nonreciprocal system, a state-of-the-art technology~\cite{shitrit2018asymmetric}. Instead of depending on magnetic or electric or nonlinear materials, which can be constrictive or challenging to incorporate into small systems, these metasurfaces provide a means of achieving nonreciprocal behaviour, or light propagation that varies with direction. However, these metasurfaces also either effectively create asymmetric dielectric and magnetic tensors or use external active tuning to provide the nonreciprocal effect~\cite{poddar2022automatic}. For example, the phase change materials were used as a base material for the metasurface to achieve the nonreciprocal behaviour~\cite{tripathi2024nanoscale}, and the self-induced nonreciprocity due to the nonreciprocal nonlinear susceptibility of the optical medium was reported~\cite{wang2025self}. Recently, a significant nonreciprocal transmission was proposed in an asymmetrical optical nonlinear metasurface that integrates bound states in the continuous (BIC) with effective zero-index media~\cite{bisignificant}.

 Feng et al. demonstrated that a multilayer stack of absorbing and non-absorbing materials, where each unit cell consisted of four alternating thin-film layers of silicon and silica on a glass substrate, shows unidirectional reflections at exceptional points, and the results were analysed by the transfer-matrix method (TMM)~\cite{feng2013demonstration}. Though the system was planar, it required nanometer-level precision in the thicknesses of the thin films. Shen et al. proposed by inverse calculation using TMM that an optical system of two layers in air shows unidirectional invisibility if the constitutive materials are lossy~\cite{shen2014unidirectional}. Additionally, when the lossy parts of the refractive indices are adjusted to their odd symmetric forms, the unidirectional invisibility can be transformed into unidirectional reflection. In the latter two examples, the transmissions were taken to be the same for the forward and backward propagations. A general concept for producing nonreciprocity by exploiting energy loss in the optical system was theoretically proposed by Huang et al.~\cite{huang2021loss}. Regardless of the direction of energy transmission, a phase lag is induced by the loss in a resonance mode. Nonreciprocity arises from the interference caused by the combination of multichannel lossy resonance modes. Recently, it was proved analytically and experimentally demonstrated that an optical system made of a single layer where both sides of the medium are not the same, shows optical nonreciprocity, provided at least one material is lossy~\cite{pratap2025nonreciprocal}. Transmission will always be nonreciprocal, except in the case of a normal incident when both sides of the media are lossless. Moreover, the value of nonreciprocity is small. However, using the linear isotropic dielectric materials, without any external biasing, and only based on optical scattering, a strong nonreciprocal system has not been reported yet. 

In this study, we present a strong nonreciprocal effect on a single layer of structured surfaces that relies only on asymmetric Mie scattering; no extrinsic biasing, electromagnetic coupling, or nonlinear effect are utilized to produce this effect. In order to show such a strong nonreciprocity in that optical system, we only use linear non-magnetic dielectric Lorentz reciprocal materials.   

\section{Results}
\begin{figure}[h]
\centering
\includegraphics[width = 1\textwidth]{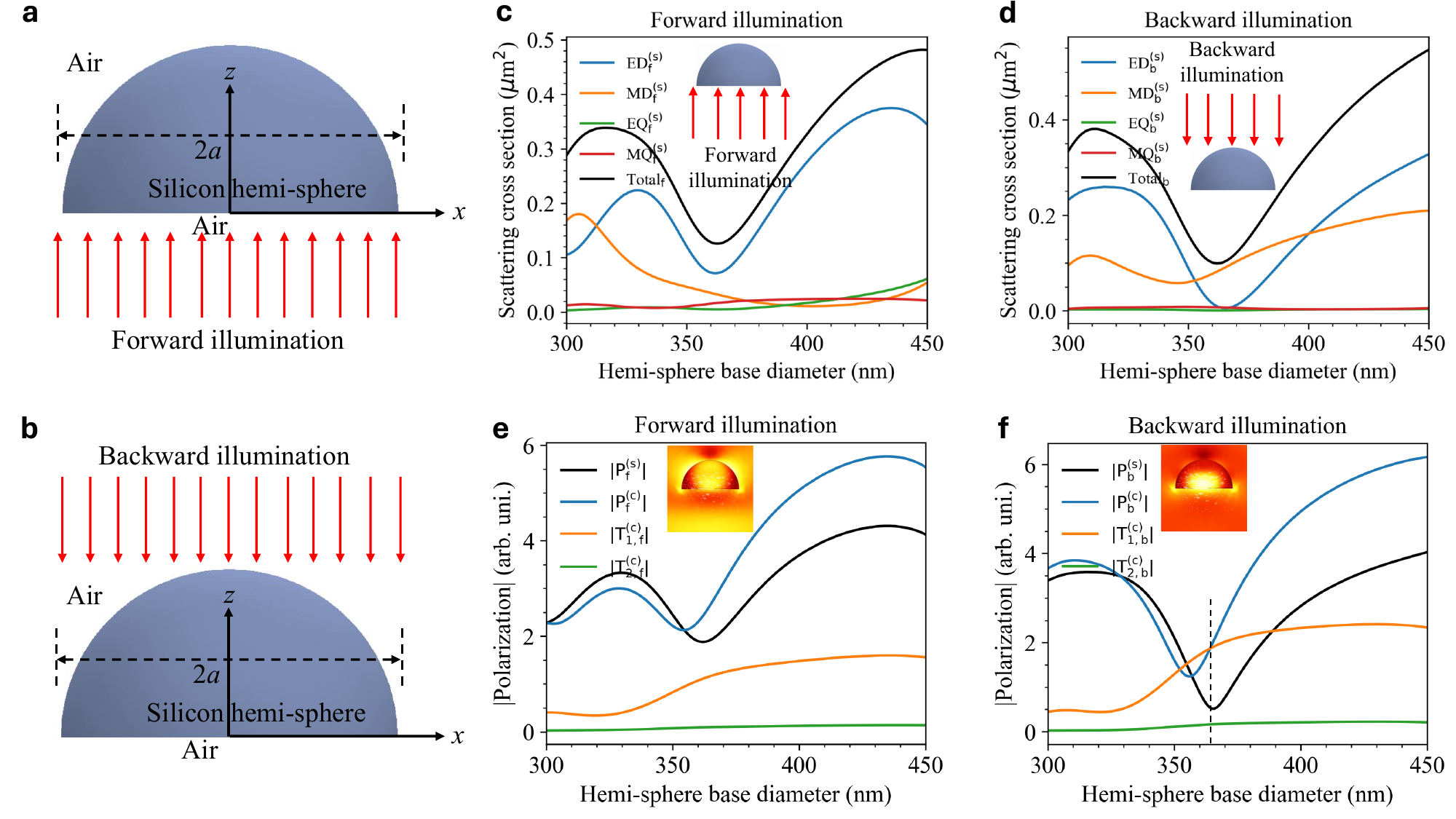}
\caption{\textbf{Asymmetric scattering and anapole}. \textbf{a, b}, Schematic of forward, along the $+z$-direction, and backward, along $-z$-direction, illuminations of hemispherical amorphous silicon nanoparticle. \textbf{c, d}, Scattering cross section of multipoles in spherical basis for the forward and backward illuminations. \textbf{e, f}, Electric dipole norms in spherical and Cartesian bases for the forward and backward illuminations. Illuminating light wavelength is 850~nm. Inset in \textbf{e,f} shows the electric field norm (thermal color map) and electric field vectors (white arrows).}
\label{fig:Asym_anapole}
\end{figure}

A hemispherical nanoparticle gives asymmetric light scattering, and those solutions can be found numerically. Figure~\ref{fig:Asym_anapole} shows the asymmetric scattering and the existence of the anapole mode in a hemispherical nanoparticle of amorphous silicon. Figure~\ref{fig:Asym_anapole}a shows the forward illumination, along the $+z$ (base-to-apex), of the hemispherical nanoparticle of base diameter $p = 2a$, where $a$ is the radius of the same. The surrounding medium of the hemisphere is air. Figure~\ref{fig:Asym_anapole}b shows the backward illumination, along $-z$ (apex-to-base), of the same hemispherical nanoparticle. The base of the hemisphere lies in the $ xy$-plane. This convention of forward and backward propagation is not unique, and it can be chosen either way. For the forward illumination, the calculated scattering cross sections of the multipoles in spherical basis, electric dipole (ED$^{(\mathrm{s})}_\mathrm{f}$), magnetic dipole (MD$^{(\mathrm{s})}_\mathrm{f}$), electric quadrupole (EQ$^{(\mathrm{s})}_\mathrm{f}$), magnetic quadrupole (MQ$^{(\mathrm{s})}_\mathrm{f}$), are shown in Fig.~\ref{fig:Asym_anapole}c. Similarly, for the backward illuminations, the calculated scattering cross sections of the multipoles in spherical basis, electric dipole (ED$^{(\mathrm{s})}_\mathrm{b}$), magnetic dipole (MD$^{(\mathrm{s})}_\mathrm{b}$), electric quadrupole (EQ$^{(\mathrm{s})}_\mathrm{b}$), magnetic quadrupole (MQ$^{(\mathrm{s})}_\mathrm{b}$), are shown in Fig.~\ref{fig:Asym_anapole}d. In the current work, the other higher-order multipoles have been neglected since their contributions are relatively small compared to lower orders. Since individual multipoles have asymmetric character for the forward and backward illuminations, their total scattering cross section also does. 
The multipole moments in the spherical basis can be presented as a sum of more primitive components in the Cartesian basis~\cite{miroshnichenko2015nonradiating,barati2024non}
\begin{align}
\bm{p}^{\mathrm{(s,e)}}_\mathrm{f,b}(\omega) & = \bm{p}^{\mathrm{(c,e)}}_\mathrm{f,b}(\omega)+\frac{i k_0 \epsilon_d}{c}\bm{T}^{\mathrm{(c,e)}}_\mathrm{1,f,b}+\frac{i k^3_0 \epsilon^2_d}{c}\bm{T}^{\mathrm{(c,e)}}_\mathrm{2,f,b}, \nonumber \\
\bm{p}^{\mathrm{(s,m)}}_\mathrm{f,b}(\omega) & = \bm{p}^{\mathrm{(c,m)}}_\mathrm{f,b}(\omega)+\frac{i k_0 \epsilon_d}{c}\bm{T}^{\mathrm{(c,m)}}_\mathrm{f,b}, \nonumber \\
\hat{Q}^{\mathrm{(s,e)}}_\mathrm{f,b}(\omega) & = \hat{Q}^{\mathrm{(c,e)}}_\mathrm{f,b}(\omega)+\frac{i k_0 \epsilon_d}{c}\hat{T}^{\mathrm{(c,e)}}_\mathrm{f,b}, \nonumber \\
\hat{Q}^{\mathrm{(s,m)}}_\mathrm{f,b}(\omega) & = \hat{Q}^{\mathrm{(c,m)}}_\mathrm{f,b}(\omega)+\frac{i k_0 \epsilon_d}{c}\hat{T}^{\mathrm{(c,m)}}_\mathrm{f,b},
\end{align} 
where super scripts `$^\mathrm{s}$', `$^\mathrm{c}$', `$^\mathrm{e}$', and `$^\mathrm{m}$' correspond to spherical and Cartesian bases, electric and magnetic multipoles, subscripts `$_\mathrm{f}$' and `$_\mathrm{b}$' correspond to forward and backward illuminations. The symbols $\bm{p}$, $\bm{T}$, $\hat{Q}$, and $\hat{T}$ represent dipole, toroidal dipole, quadrupole, and toroidal quadrupole. Subscript indices `$_\mathrm{1}$' and `$_\mathrm{2}$' correspond to type~I and type~II toroidal dipoles. Figure~\ref{fig:Asym_anapole}e and Figure~\ref{fig:Asym_anapole}f show the amplitude of the electric polarization in Spherical basis and its corresponding Cartesian components for the forward and backward illuminations, respectively. We see that in the Cartesian basis, only the electric dipole is dominating, the toroidal dipole of type~I is significant, and the toroidal dipole of type~II contribution is negligible. Figure~\ref{fig:Asym_anapole}e shows that there is no interference situation between $\bm{p}^{\mathrm{(c,e)}}_\mathrm{f}(\omega)$ and $\bm{T}^{\mathrm{(c,e)}}_\mathrm{1,f}$. While in Fig.~\ref{fig:Asym_anapole}b, there is interference between  $\bm{p}^{\mathrm{(c,e)}}_\mathrm{b}(\omega)$ and $\bm{T}^{\mathrm{(c,e)}}_\mathrm{1,b}$, and that is the location of the anapole mode. The vertical dashed line shows the destructive interference position. At this destructive interference position, the electric field is trapped inside the nano scatterer. The insets of Fig.~\ref{fig:Asym_anapole}e and Fig.~\ref{fig:Asym_anapole}f show the electric field distribution in the hemispherical silicon nanoparticle. We see that the anapole exists only for backward illumination. Due to the anapole, the electric field will be trapped inside and will not allow it to propagate in the far-field regime. 

\begin{figure}[h]
\centering
\includegraphics[width = 1\textwidth]{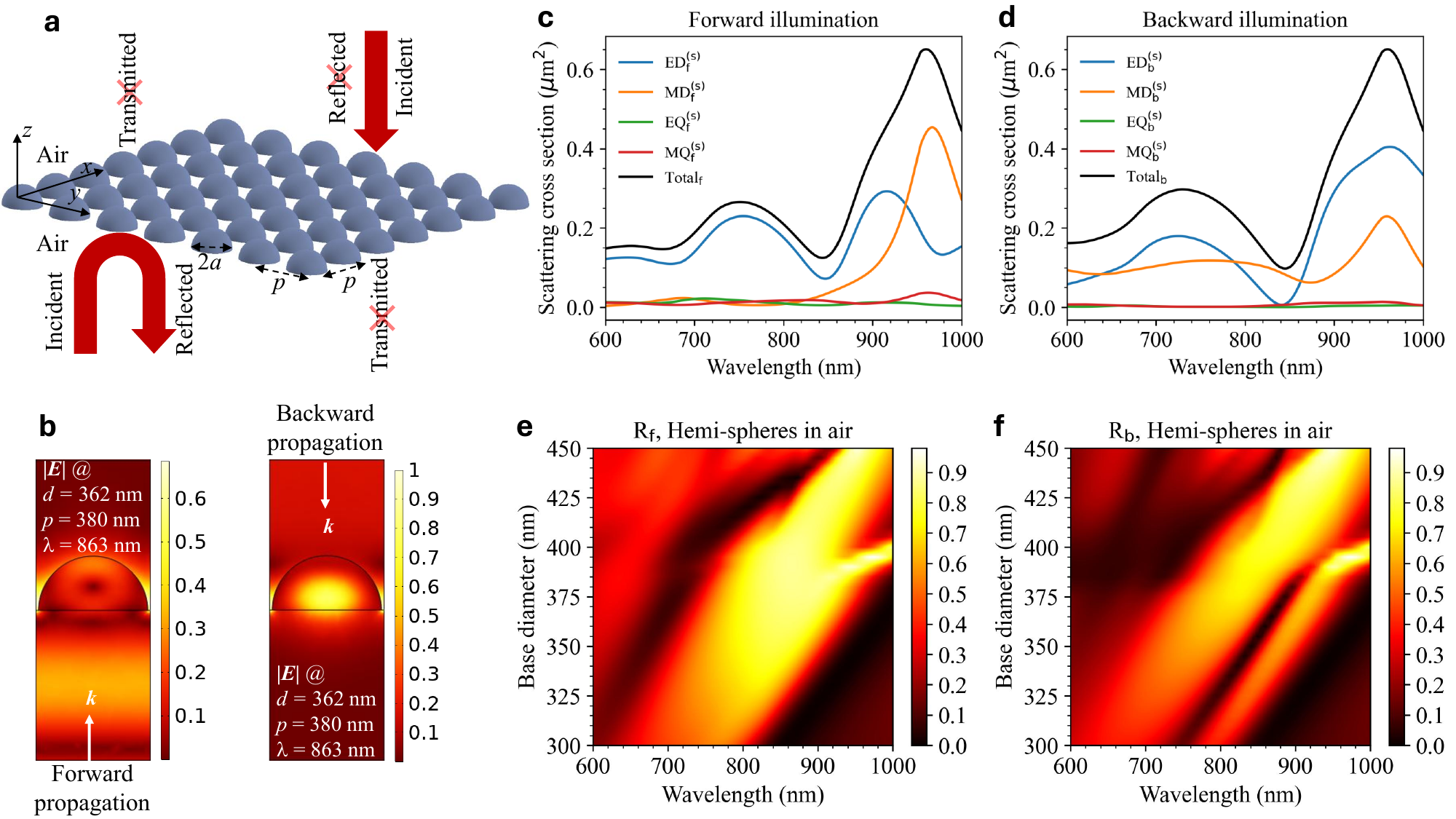}
\caption{\textbf{Nonreciprocal Mie-surface in air}. \textbf{a,} Schematic presentation of the nonreciprocal behaviour of the Mie-surface placed in air with forward and backward illuminations. \textbf{b,} Thermal color map of electric field norms for the backward and forward illuminations at wavelength $\lambda = 863$~nm, period $p = 380$~nm, and  hemisphere base diameter $d (2a) = 362$~nm. \textbf{c, d,} Scattering cross section of multipoles for the forward and backward illuminations in the spherical basis at fixed period $p = 380$~nm and base diameter $d = 362$~nm. \textbf{e, f,} Thermal color map reflectance for the forward and backward illuminations at period $p = 380$~nm.}
\label{fig:Scat_in_air}
\end{figure}
Arranging the discussed sub-wavelength scale hemispheres in a rectangular grid in the $xy$-plane, we get a nanotructured surface with a height equal to the radius of these scatterers. Figure~\ref{fig:Scat_in_air}a shows a schematic of such a surface made of hemispheres. Its optical properties for forward and backward propagation waves depend purely on the Mie-scattering modes. Therefore, we termed these sub-wavelength structured surfaces as Mie-surfaces. The schematic shows the interaction of the Mie-surface with forward and backward propagating waves. Figure~\ref{fig:Scat_in_air}b shows the electric field norm in a unit cell of the Mie-surface at wavelength 863~nm, hemisphere base diameter 362~nm, and periodicity 380~nm for the forward and backward interactions. It can be seen that for the backward propagation, the electric field is trapped inside the hemispherical nanoparticle. The scattering cross section of multipoles in spherical basis are shown in Fig.~\ref{fig:Scat_in_air}c for forward illumination and Fig.~\ref{fig:Scat_in_air}d for backward illumination. We see that scattering is mainly dominated by only electric and magnetic dipoles (ED$^\mathrm{(s)}_\mathrm{f,b}$, MD$^\mathrm{(s)}_\mathrm{f,b}$) and higher order multipoles, electric quadrupole and magnetic quadrupole (EQ$^\mathrm{(s)}_\mathrm{f,b}$, MD$^\mathrm{(s)}_\mathrm{f,b}$),  are negligible. In the far-field domain, the calculated density plot of reflectance (R$_\mathrm{f,b}$) with respect to the wavelength and base diameter of hemispheres are shown in Fig.~\ref{fig:Scat_in_air}e and Fig.~\ref{fig:Scat_in_air}f at the normal incident of light for forward and backward illuminations, respectively. We see that in the spatio-spectral regime where the anapole mode occur, have maximum differences between the both reflectance. In other words, maximum nonreciprocity in reflection occurs in that spatio-spectral regime. At the normal incident, transmission does not have any nonreciprocity (Supplementary information). Transmission nonreciprocity of a slab will only occur at the normal incident when the media on both sides of the Mie-surface optical system are asymmetric, and at least one medium is lossy as given by the T-function~\cite {pratap2025nonreciprocal},
\begin{align}
F_\mathrm{T} (n_1,n_2,n_3) = \frac{16 n_2 n^*_2 (n^*_1 n_3 - n_1 n^*_3)}{(n_1+n_2)(n^*_1+n^*_2)(n_2+n_3)(n^*_2+n^*_3)},
\label{eq:FT}
\end{align}
where $n_1$, $n_2$, and $n_3$ are the complex refractive indices of the first medium (upper), second medium (slab), and third medium (lower or substrate), and symbol `$^{*}$' denotes the complex conjugate. However, at the non-zero incident angle of the light, transmission nonreciprocity will be achieved even if the same medium is on both sides of the Mie-surface because the Mie-surface slab works as effectively an inhomogeneous medium.  

\begin{figure}[h]
\centering
\includegraphics[width = 1\textwidth]{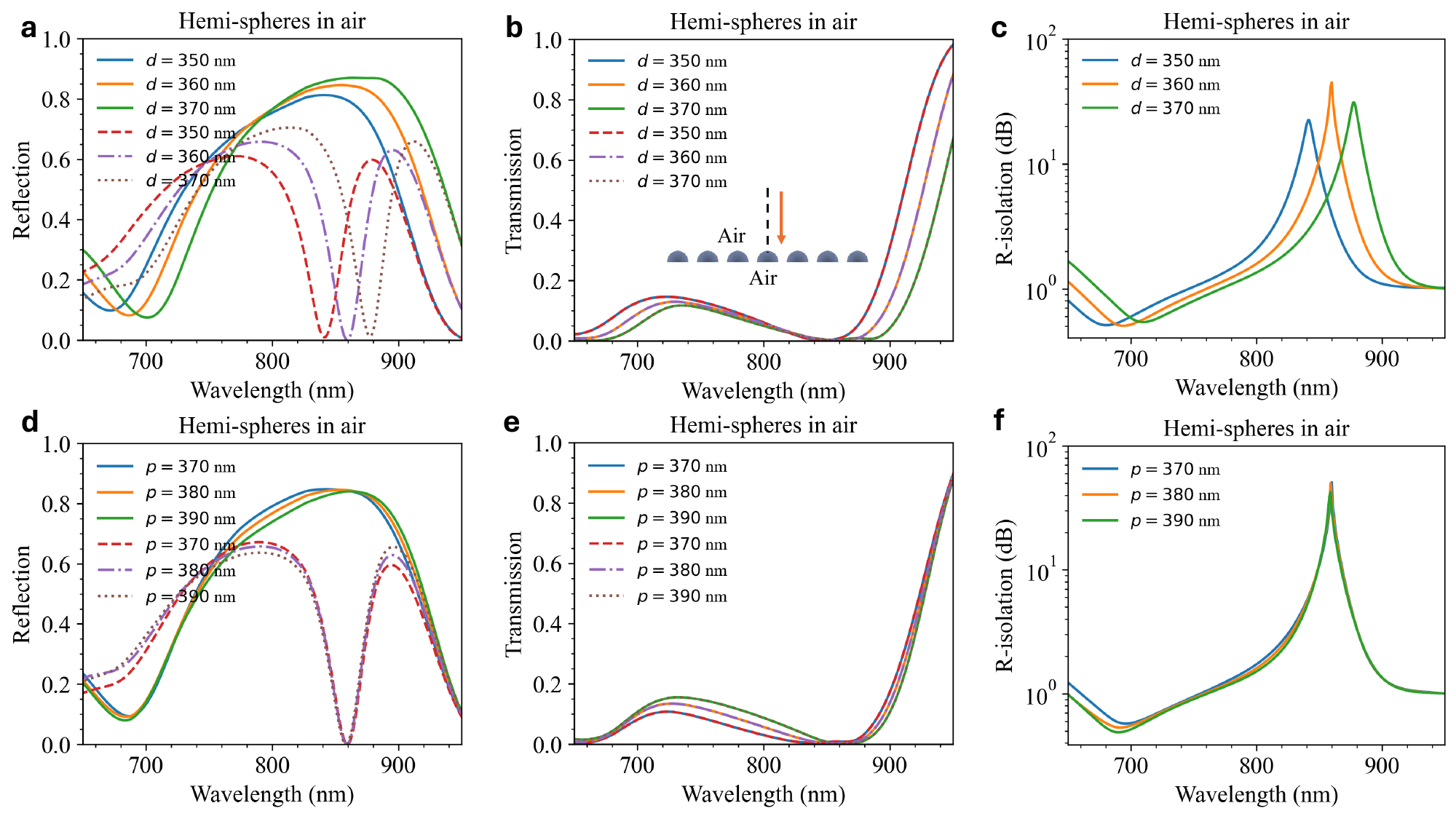}
\caption{\textbf{Nonreciprocity of Mie-surface in air.} Top panel: \textbf{a,} Reflectance, \textbf{b,} transmittance,  and \textbf{c,} reflection(R)-isolation ratio for the forward and backward illuminations at fixed period $p = 380$~nm and different base diameters $d=$ 350~nm, 360~nm, and 370~nm. Bottom panel: \textbf{d,} Reflectance, \textbf{e,} transmittance, and \textbf{f,} R-isolation for the forward and backward illuminations at fixed diameter $d = 360$~nm and different periods $p=$ 370~nm, 380~nm, and 390~nm. } 
\label{fig:RnT_in_air}
\end{figure}
The light propagation characteristics of the Mie-surface are very interesting, due to the anapole mode condition that only depends on the size of the scatterer. Figure~\ref{fig:RnT_in_air} shows the reflection, transmission, and isolation characteristics of the Mie-surface. First, we fixed the period ($p=380$~nm) of the Mie-surface and varied the base diameter of the hemisphere. Figure~\ref{fig:RnT_in_air}a shows the forward (continuous lines) and backward (dashed lines) reflections at different base diameters of the hemisphere, 350~nm, 360~nm, and 370~nm. We see that the peak in forward reflection and dip in backward reflection significantly shift with changing base diameter of the hemisphere. Transmission in both directions is the same as shown in Fig.~\ref{fig:RnT_in_air}b at normal incidence due to a lossless medium on both sides of the Mie-surface. The relative contrast between the forward and backward reflection, R-isolation, is shown in Fig.~\ref{fig:RnT_in_air}c. The peak values of these R-isolations are 22~dB, 45~dB, and 31~dB at wavelength 842~nm, 860~nm, and 878~nm, respectively. Now, we fixed the base diameter 360~nm of the hemispherical nanoparticle, and changed the periodicity to 370~nm, 380~nm, and 390~nm. Figure~\ref{fig:RnT_in_air}d shows the forward (continuous lines) and backward (dashed lines) reflection. We see that the periodicity does not affect the peaks and dips' spectral position significantly. The reason is the existence of the anapole that only depends on the size of the nanoparticle, not on the periodicity. Transmission characteristic in Fig.~\ref{fig:RnT_in_air}e is similar to the former one as both sides of the Mie-surface medium are lossless. The relative contrast between the reflectance, R-isolation, at different periodicity is shown in Fig.~\ref{fig:RnT_in_air}f. The isolation peaks have values 51~dB, 50~dB, and 42~dB at wavelength 860~nm, 859~nm, and 859~nm, respectively. We see that the periodicity effect on the nonreciprocity of the Mie-surface is negligible. 

\begin{figure}[h]
\centering
\includegraphics[width = 1\textwidth]{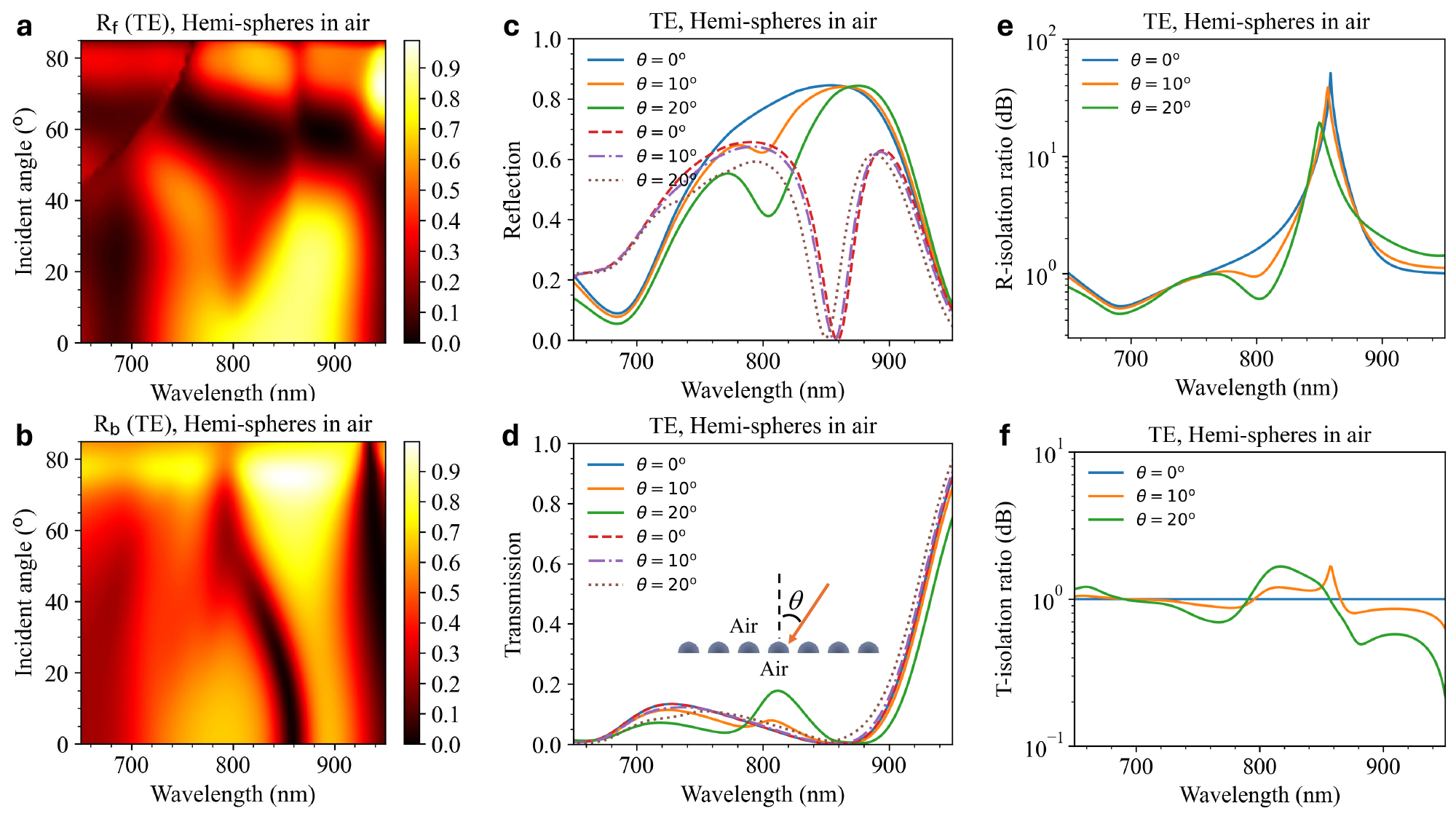}
\caption{\textbf{Nonreciprocity of Mie-surface in air with TE polarized light.} \textbf{a, b,} Density map of reflectance with incident angle and wavelength for transverse electric (TE) polarization of light. \textbf{c, d} Reflection and transmission for TE polarization at incident angle $\theta=0^{\circ}, 10^{\circ}, 20^{\circ}$. \textbf{e, f,} Isolation ratios for reflection and transmission. In \textbf{a-f}, the period $p = 380$~nm and diameter $d = 360$~nm are fixed.}
\label{fig:RnT_in_air_TE}
\end{figure}
The light polarizations, transverse electric (TE) and transverse magnetic (TM), also have some different effects on the nonreciprocity of the Mie-surface. Figure~\ref{fig:RnT_in_air_TE} shows the TE polarization effect on the nonreciprocity. Figure~\ref{fig:RnT_in_air_TE}a and Fig.~\ref{fig:RnT_in_air_TE}b show the density plot of the forward (R$_\mathrm{f}$) and backward (R$_\mathrm{b}$) reflections, respectively, with varying wavelength and incident angle of the TE polarized light. We see that there is a huge difference between the forward and backward reflections. The forward (T$_\mathrm{f}$) and backward (T$_\mathrm{b}$) transmission also have significantly different values from each other for polarized light (Supplementary information). Figure~\ref{fig:RnT_in_air_TE}c shows the forward (continuous lines) and backward (dashed lines) reflections together at three different incident angles $\ theta=$0$^ {\circ}$, 20$^{\circ}$, and 20$^{\circ}$ of the TE polarization. Figure~\ref{fig:RnT_in_air_TE}d shows the transmission for the forward (continuous lines) and backward (dashed lines) illuminations at the same three different incident angles as are for reflections. We see that at zero incident angle, both directions of transmission are the same, while for any non-zero incident angle, they are different. However, this difference in transmission is not as much as compared to reflection. Figure~\ref{fig:RnT_in_air_TE}e and Fig.~\ref{fig:RnT_in_air_TE}f show the relative contrast reflection R-isolation and transmission T-isolation, respectively, for those three incident angles in Fig.~\ref{fig:RnT_in_air_TE}c and Fig.~\ref{fig:RnT_in_air_TE}d. The R-isolation peaks have values of 51~dB, 38~dB, and 19~dB, at wavelength 859~nm, 857~nm, and 850~nm. The T-isolation for zero-incident is flat as both transmissions are the same. While T-isolation for non-zero incident angle does not have any interesting contrast, and it is less than 10~dB.  

\begin{figure}[h]
\centering
\includegraphics[width = 1\textwidth]{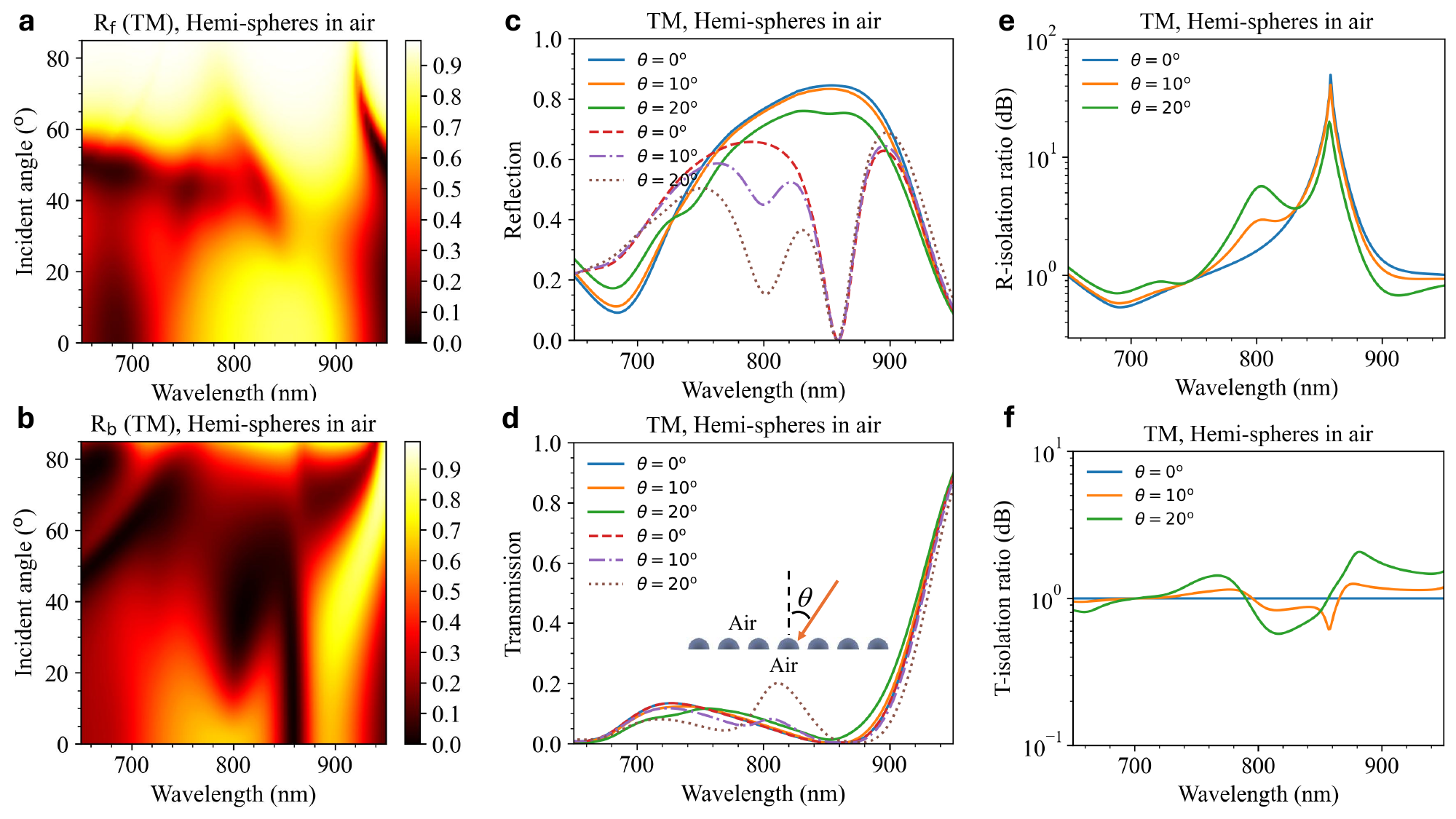}
\caption{\textbf{Nonreciprocity of Mie-surface in air with TM polarized light.} \textbf{a, b,} Density map of reflectance with incident angle and wavelength for transverse electric (TM) polarization of light. \textbf{c, d} Reflection and transmission for TM polarization at incident angle $\theta=0^{\circ}, 10^{\circ}, 20^{\circ}$. \textbf{e, f,} Isolation ratios for reflection and transmission. In \textbf{a-f}, period $p = 380$~nm and diameter $d = 360$~nm are fixed.}
\label{fig:RnT_in_air_TM}
\end{figure}
The effect of TM polarization on the nonreciprocity of the Mie-surface is shown in Fig.~\ref{fig:RnT_in_air_TM}. The density plot of the forward (R$_\mathrm{f}$) and backward (R$_\mathrm{b}$) reflection with respect to the incident angle and wavelength is shown in Fig.~\ref{fig:RnT_in_air_TM}a and Fig.~\ref{fig:RnT_in_air_TM}b, respectively. We see that there is a huge distinction between the two reflections. As the incident angle reaches the grazing angle, the maximum light is reflected back at any wavelength for the forward propagation, while this does not happen in the backward propagation due to the curved surface of the nanoparticle. The transmission density plots in the case of TM polarization for the forward (T$_\mathrm{f}$) and backward (T$_\mathrm{b}$) propagations for TM polarizations have less contrast than reflections (Supplementary information). Figure~\ref{fig:RnT_in_air_TM}c shows the forward (continuous lines) and backward (dashed lines) reflection together at a few incident angles $\theta=0^{\circ}, 10^{\circ}, 20^{\circ}$, for relative comparison. Here, at lower incident angles, significant secondary minima are observed for backward reflection, and not for forward reflection. Figure~\ref{fig:RnT_in_air_TM}d shows the forward and backward transmission. We see that for the non-zero incident angles, forward (continuous lines) and backward (dashed lines) transmissions are slightly different, and this difference increases with increasing incident angle. Figure~\ref{fig:RnT_in_air_TM}e and Fig.~\ref{fig:RnT_in_air_TM}f show the R-isolation and T-isolation for a few incident angles as are in Fig.\ref{fig:RnT_in_air_TM}c and Fig.~\ref{fig:RnT_in_air_TM}d. The peak values of R-isolation are 50~dB, 41~dB, and 20~dB, at spectral positions 859~nm, 859~nm, and 858~nm, respectively. These peak positions are very close compared to the case of TE polarization. For the TM polarization case also, the T-isolation is very small, around 10~dB, and it is similar to the TE polarization.  

\begin{figure}[h]
\centering
\includegraphics[width = 1\textwidth]{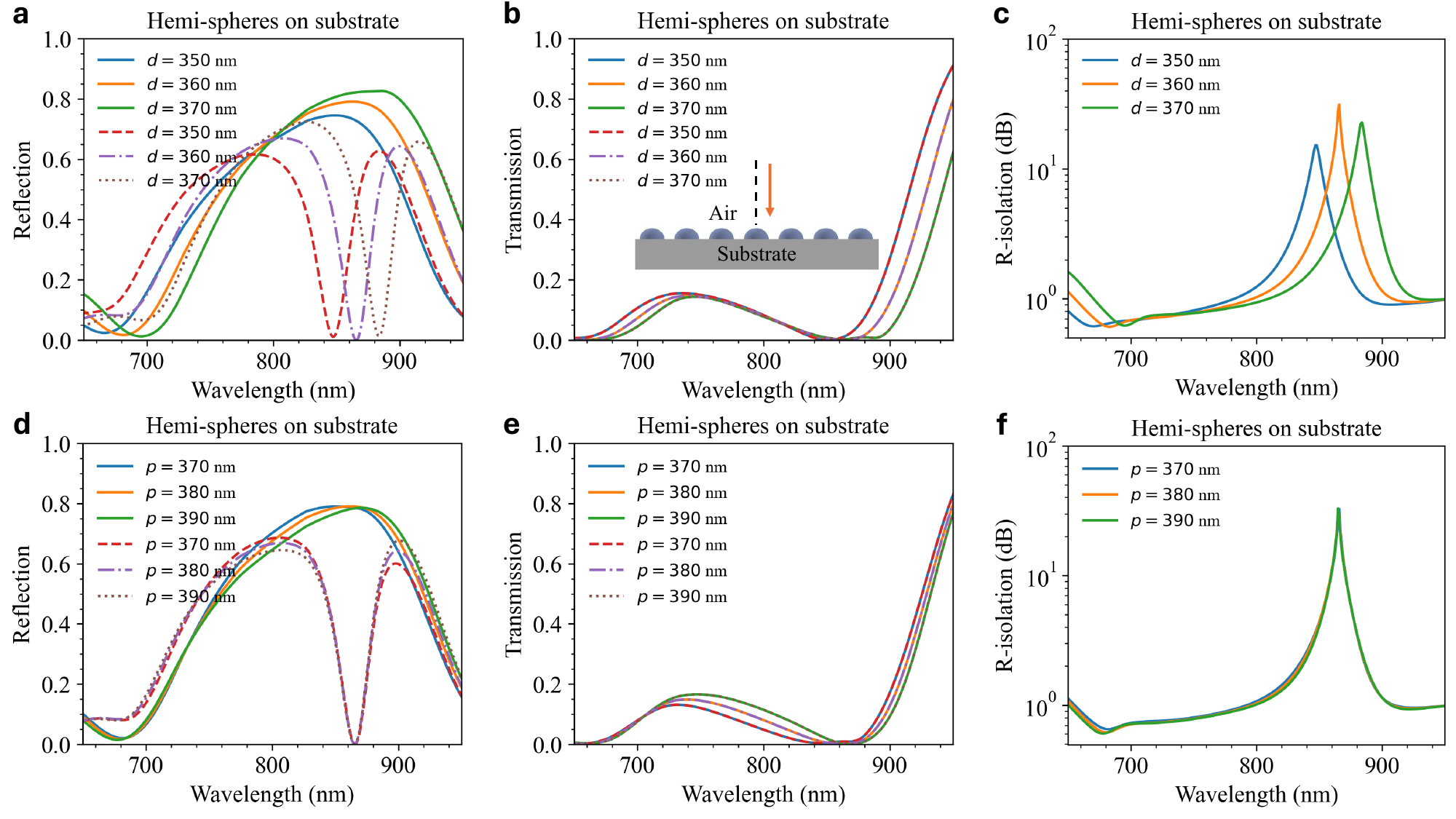}
\caption{\textbf{Nonreciprocity of Mie-surface on substrate.} Top panel: \textbf{a,} Reflectance, \textbf{b,} transmittance and \textbf{c,} reflection isolation ratio for the forward and backward propagation at fixed period $p = 380$~nm at different hemisphere base diameters 350~nm, 360~nm, and 370~nm. Bottom panel:  \textbf{d,} Reflectance, \textbf{e,} transmittance and \textbf{f,} transmission isolation ratio for the forward and backward propagation at fixed base diameter $d = 360$~nm at different periodicity 370~nm, 380~nm, and 390~nm.}
\label{fig:RnT_on_subs}
\end{figure}
In all the previously considered cases, the Mie-surfaces were embedded in air, i.e, on both sides, the mediums were the same. In general, experimentally, the Mie-surfaces will be on some substrate. Further, we explore the substrate effect on the light propagation characteristic of the Mie-surfaces compared to the case when these were embedded in air. Figure~\ref{fig:RnT_on_subs} shows the effect of substrate (lossless glass) on the nonreciprocity of the Mie-surfaces at the normal incident of the light wave. In Fig.~\ref{fig:RnT_on_subs}a, Fig.~\ref{fig:RnT_on_subs}b, and Fig.~\ref{fig:RnT_on_subs}c, the periodicity is fixed at 380~nm, and there are variable base diameter of the hemisphere at 350~nm, 360~nm, and 370~nm. Fig.~\ref{fig:RnT_on_subs}a shows the forward (continuous lines) and backward (dashed lines) reflections. We observe a red-shift of nearly 6~nm compared to the case of Fig.~\ref{fig:RnT_in_air}a. Figure~\ref{fig:RnT_on_subs}b shows the forward and backward transmission of the Mie-surfaces on the substrate. Here, the transmission for both illuminations is the same due to the lossless medium on both sides of the Mi-surface as predicted by the T-function of Eq.~(\ref{eq:FT}). Figure~\ref{fig:RnT_on_subs}c shows the R-isolation. The peak values of R-isolation 15~dB, 31~dB, and 23~dB, at wavelength 847~nm, 866~nm, and 884~nm, respectively. We observe a reduction in the R-isolation and a red shift in the spectral position of these peaks compared to Fig.~\ref{fig:RnT_in_air}c—for example, the highest peak reduction of 14~dB, and a red shift of 6~nm. Now, we fixed the base diameter of the hemisphere to 360~nm, and varied the periodicity to 370~nm, 380~nm, and 390~nm. Figure~\ref{fig:RnT_on_subs}d shows the forward and backward reflection. We see that there is no peak or dip position variation at different periodicities. Figure~\ref{fig:RnT_on_subs}e shows transmission for both types of illumination. The forward and backward transmissions are the same due to lossless media on both sides of the Mie-surfaces. Figure~\ref{fig:RnT_on_subs}f shows the R-isolation. The peak numeric values of the R-isolation are 33~dB, 31~dB, and 33~dB, corresponding to spectral positions at 866~nm, 865~nm, and 865~nm, respectively. It is clear that the variation in the periodicity does not affect the R-isolation, and it remains almost undisturbed. However, there is R-isolation reduction of 19~dB, and a 6~nm spectral red-shift for the periodicity 380~nm as compared to the case of Fig.~\ref{fig:RnT_in_air}f.  

\begin{figure}[h]
\centering
\includegraphics[width = 1\textwidth]{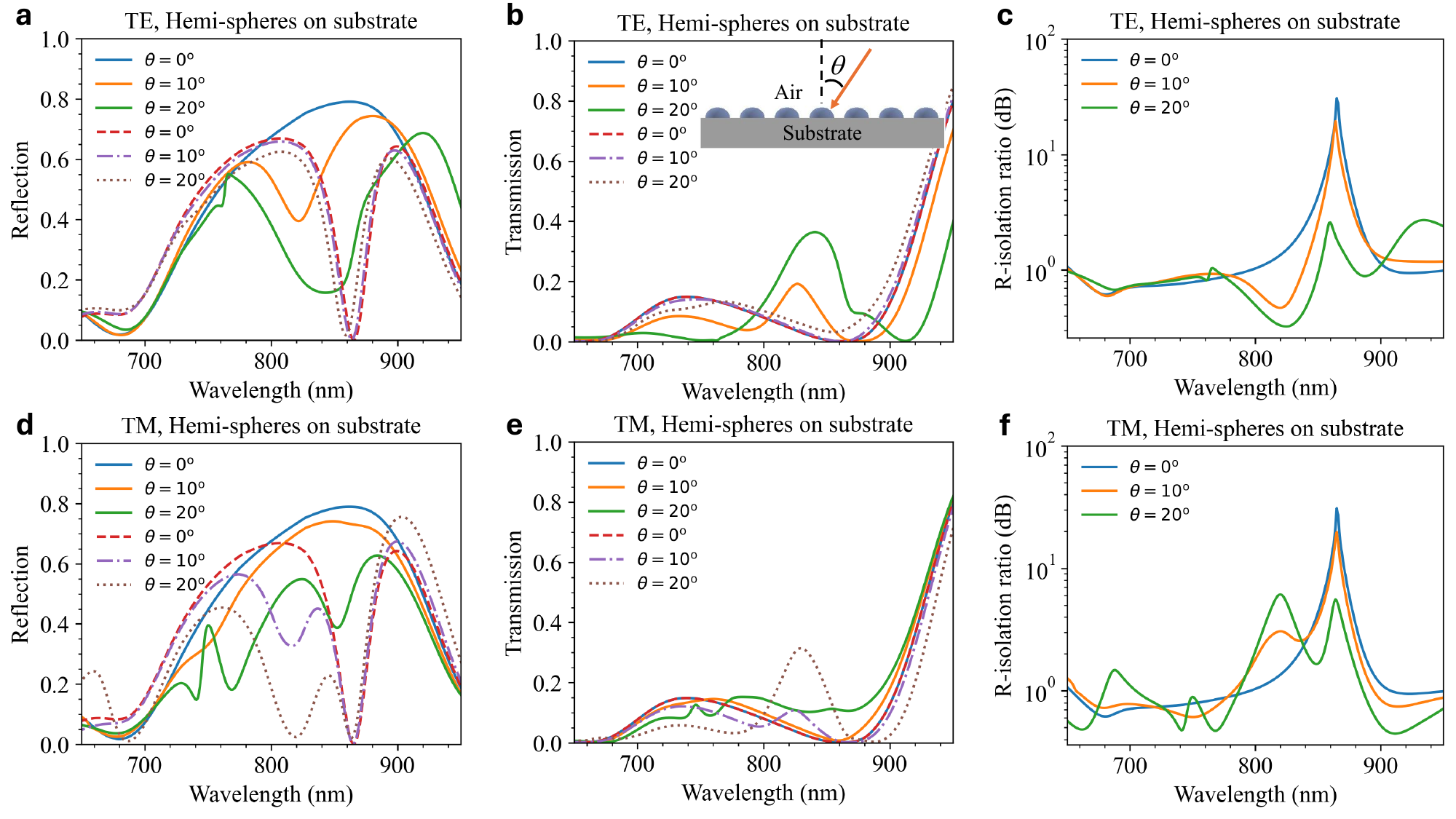}
\caption{\textbf{Nonreciprocity of Mie-surface on substrate for polarized light.} Top panel: \textbf{a,} Reflectance, \textbf{b,} transmittance, and \textbf{c,} reflection isolation ratios for forward and backward illuminations at different indecent angle of the transverse electric (TE) polarized light. Bottom panel: \textbf{d,} Reflectance, \textbf{e,} transmittance, and \textbf{f,} reflection isolation ration for forward and backward illuminations at different indecent angle of the transverse magnetic (TM) polarized light. Here, period $p = 380$~nm, and diameter $d = 360$~nm are fixed.}
\label{fig:RnT_on_subs_TE_TM}
\end{figure}
The effect of substrate on the properties of Mie-surfaces interacting with TE and TM polarized light is summarized in Fig.~\ref{fig:RnT_on_subs_TE_TM}. Figure~\ref{fig:RnT_on_subs_TE_TM}a-c corresponds to TE polarization and Fig.~\ref{fig:RnT_on_subs_TE_TM}d-f corresponds to TM polarization of light at the indecent angles $\theta=0^{\circ}$, $\theta=10^{\circ}$, and $\theta=20^{\circ}$. The reflection, transmission, and isolation for the forward (continuous lines) and backward (dashed lines) illuminations are shown in Fig.~\ref{fig:RnT_on_subs_TE_TM}a, Fig.~\ref{fig:RnT_on_subs_TE_TM}b, and Fig.~\ref{fig:RnT_on_subs_TE_TM}c, respectively. Due to the substrate, as the angle increases, there is a faster change in the forward reflection, while the backward reflection is not affected much. For the non-zero incident, transmission for both illuminations is significantly different. The R-isolation has well-defined maxima peaks at 865~nm, 864~nm, and 860~nm, and their corresponding values are 31~dB, 19~dB, and 3~dB, respectively. Figure~\ref{fig:RnT_on_subs_TE_TM}d shows the forward and backward reflection at a few incident angles. Backward reflections change faster than forward, and there are secondary minima to the left that adjust to the global minima. Figure~\ref{fig:RnT_on_subs_TE_TM}e shows the transmission for the forward and backward propagations. The relative contrast in transmission is small. Figure~\ref{fig:RnT_on_subs_TE_TM}f shows the relative reflection contrast, R-isolation, for TM polarization. The peak values of these R-isolation are 31~dB, 20~dB, and 6~dB, at wavelength 865~nm, 865~nm, and 864~nm, respectively. Other local maxima left adjacent to the peak of 864~nm are due to the local minima in the backward propagation at 20$^{\circ}$ incident angle. 

Overall, we found that the nonreciprocity of Mie-surface is very strong in the reflection. The transmission also has a minute nonreciprocity only for the non-zero incident of light due to the lossless mediums on both sides. The substrate reduces the isolation ratio of Mie-surface slightly compared to when it is embedded in air. However, the isolation reduction can be minimized by reducing the refractive index contrast between both sides of the Mie-surface. In all the above analyses, we notice that the ratio of the radius of the hemisphere to the wavelength of the regime where a strong nonreciprocity occurs is roughly one-fifth. Such a single layer of Mie-surface that has a small effective thickness is sufficient to provide strong optical nonreciprocity. It should be noted that the constitutive materials, silicon and glass, are non-magnetic and taken under a linear regime; moreover, individually they are Lorentz reciprocal. The Mie-surface nonreciprocity is absolutely based on the anapole mode existence only for the propagation along the backward direction, and non-existence for the opposite direction propagation. The light propagation characteristic of the presented Mie-surface is passive, and free from any external bias. The Mie-surface can be used for passive linear nonreciprocal devices. Any other asymmetric design of a high refractive nanoparticle may give nonreciprocity if the asymmetric interference exists between the moments of the dominating multipoles.  

\section{Conclusion}
The hemispherical shape of the amorphous silicon nanoparticle in air gave the asymmetric scattering for the forward and backward direction of illuminations. A destructive interference occurred between the Cartesian electric dipole and toroidal electric dipole, leads to anapole, only for the backward illumination of the hemispherical nanoparticle. The asymmetric existence of anapole of hemispherical nanoparticle was utilised to create a structured surface.  The surface was obtained by arranging of hemispherical silicon nanoparticles in square lattice in air, and on a lossless glass substrate. That structured surface provided optical nonreciprocity purely caused by the existence of anapole, a fundamental concept of optical Mie-scattering. Therefore, that surface was termed as the Mie-surface. The Mie-surface showed strong nonreciprocal behaviour in the reflection for the upward and downward directions of illuminations, and minute nonreciprocal behaviour in transmission for non-zero incident angle. The nonreciprocity of Mie-surface was higher when it was in air compared to when it was on glass substrate. There was a slight reduction of the isolation ratio and red-shift due to the glass substrate. Presented nonreciprocal Mie-surfaces could be used for the applications of linear and passive nonreciprocal photonic devices. 

\section{Method}
The analytical solution for scattering of the hemispherical nanoparticle are not possible. We utilised finite element method (FEM) based commercial software COMSOL Multiphysics for our analysis. For scattering, spherical 3D-model was made in the Wave Optics module of the COMSOL where outside of the particle air was considered and a perfect matched layer (PML) was applied. For the reflection and transmission, a rectangular 3D-model was made in the Wave Optics module where side wall boundaries were kept as periodic boundary conditions and bottom and top boundaries were ports as source for forward and backward illuminations. Reflections and transmission were calculated using the S-parameters. Other data analysis were carried out in Python.  
 
\section*{Acknowledgement}
D. P. thanks IIT Delhi for the NFSG fund project number MI02922G, and ANRF (SERB) for the fund project number EEQ/2023/000240. 

\section*{Disclosures}
The authors declare no conflict of interest.



\bibliography{references}   
\bibliographystyle{ieeetr}
\newpage
\newcommand{\beginsupplement}{%
        \setcounter{table}{0}
        \renewcommand{\thetable}{S\arabic{table}}%
        \setcounter{figure}{0}
        \renewcommand{\thefigure}{S\arabic{figure}}%
     }
     
\section*{\LARGE{Supplementary information}} 
\beginsupplement
\section*{Linear material data}
\begin{figure}[h]
\centering
\includegraphics[width = 0.4\textwidth]{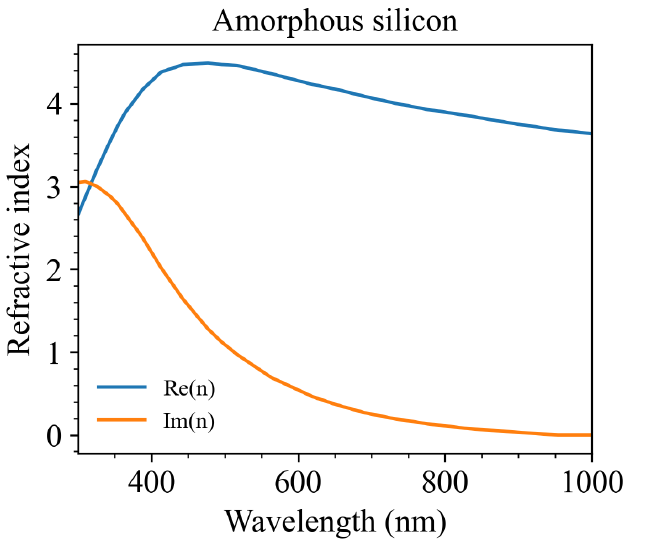}
\caption{Real and imaginary part of the refractive index of the amorphousness silicon (Si) thin film  from Ref.~\cite{pierce1972electronic} used for calculations.}
\label{fig:Si_index}
\end{figure}

In our analysis, the silicon (Si) was treated as amorphous. The real and imaginary part of the refractive of the Si are shown in Fig.~\ref{fig:Si_index} from the Ref.~\cite{pierce1972electronic}. We can see that the Si is having dispersive losses in the given spectrum range.

\section*{Transmission vs diameter vs wavelength at normal incident}
\begin{figure}[h]
\centering
\includegraphics[width = 0.8\textwidth]{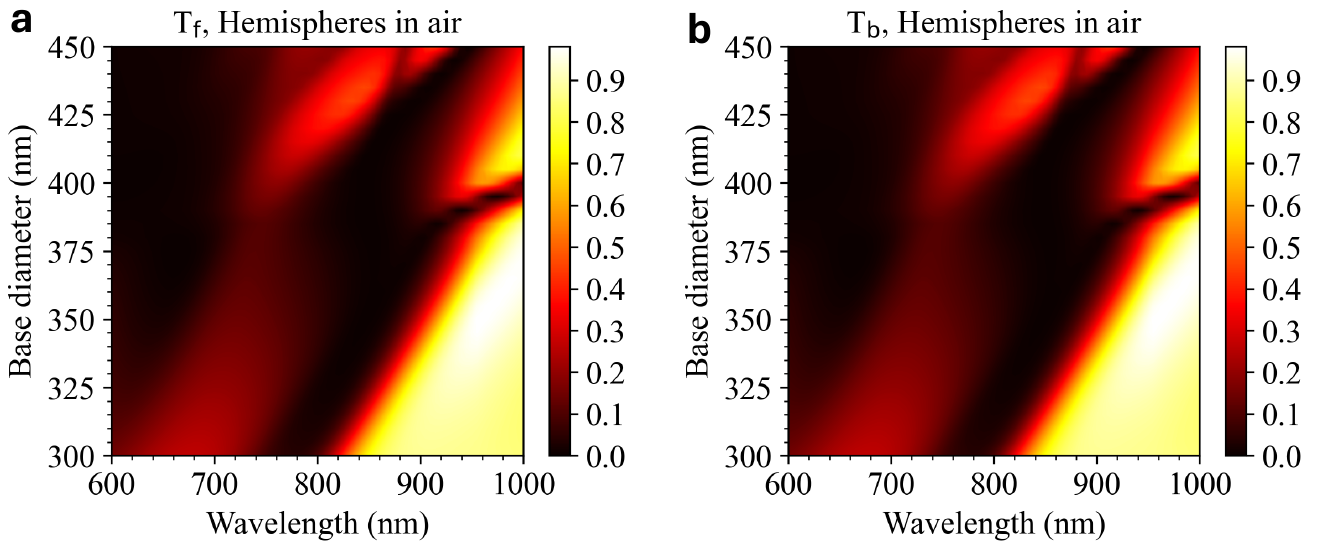}
\caption{\textbf{Reciprocal transmission of Mie-surface in air at normal incident.} \textbf{a,} Forward, and \textbf{b,} backward transmission density map with respect to diameter of the hemisphere and wavelength of the light at the normal incident and fixed period $p = 380$~nm. }
\label{fig:T_den_in_air_vs_dia_vs_lda}
\end{figure}

The transmission density plot with respect to the diameter of the hemisphere and wavelength of the light at normal incident are shown in Fig.~\ref{fig:T_den_in_air_vs_dia_vs_lda}a for the forward illumination and in Fig.~\ref{fig:T_den_in_air_vs_dia_vs_lda}b for the backward illumination, at fixed period 380~nm. A reciprocal transmission can be seen at the normal incident angle for the both illuminations. 

\section*{Transmission vs incident angle vs wavelength}
\begin{figure}[h]
\centering
\includegraphics[width = 0.8\textwidth]{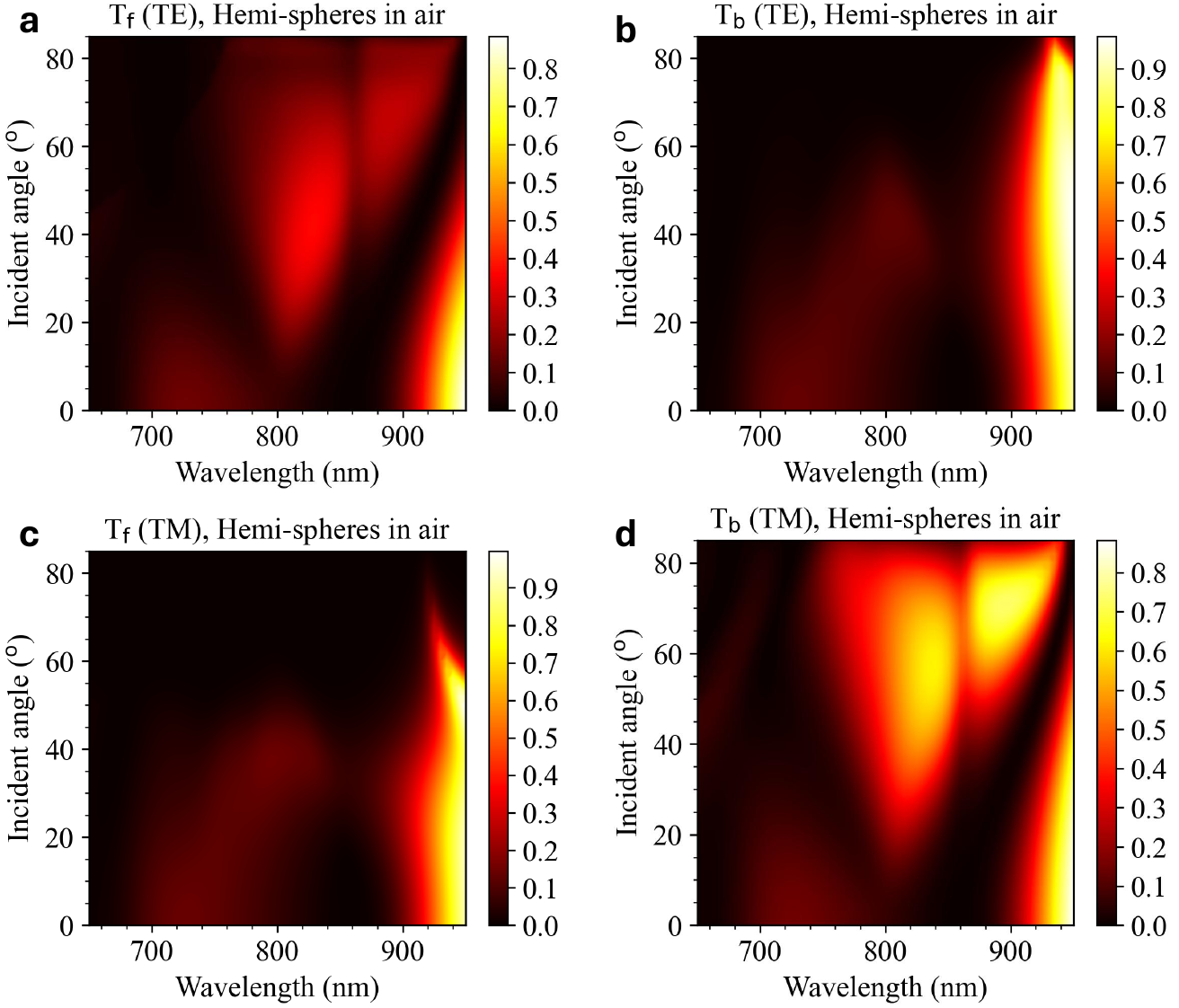}
\caption{\textbf{Nonreciprocal transmission of Mie-surface in air for polarized light.} \textbf{a-d} Forward and backward transmission density map of the for transverse electric (TE) and transverse magnetic (TM) polarizations of the incident light. The period $p = 380$~nm, diameter $d = 360$~nm are fixed. }
\label{fig:T_den_in_air_TE_TM}
\end{figure}

The transmission density plot with respect to the incident angle and wavelength of the light are shown in Fig.~\ref{fig:T_den_in_air_TE_TM}a,b, for the transverse electric (TE) polarization, and in the Fig.~\ref{fig:T_den_in_air_TE_TM}c,d, for the transverse magnetic (TM) polarization, at fixed period 380~nm and diameter 360~nm. A non-zero but small nonreciprocity can be observed in the transmission between the forward and backward illuminations.   

\end{document}